\documentclass[%
 reprint,
 amsmath,amssymb,
 aps,
]{revtex4-1}

\usepackage{graphicx}% Include figure files
\usepackage{subfigure}
\usepackage{dcolumn}% Align table columns on decimal point
\usepackage{bm}% bold math
\begin{document}

\preprint{APS/123-QED}

\title{Time and Frequency Structure of Causal Correlation Network in China Bond Market}% Force line breaks with \\

\author{Zhongxing Wang}
 \affiliation{Research Center on Fictitious Economy and Data Sciences, Chinese Academy of Sciences, Beijing, 100190, PR China}%Lines break automatically or can be forced with \\
 \affiliation{%
 School of Economics and Management, University of Chinese Academy of Sciences, Beijing, 100080, PR China\\
}%
\author{Yan Yan}
 \email{yanyan@ucas.ac.cn}

\affiliation{%
 School of Economics and Management, University of Chinese Academy of Sciences, Beijing, 100080, PR China\\
}%
\affiliation{
 Key Laboratory of Big Data Mining and Knowledge Management, Chinese Academy of Sciences,
Beijing, 100191, PR China}
\author{Xiaosong Chen}
\affiliation{
 Institute of Theoretical Physics, Chinese Academy of Sciences, Beijing, 100190, PR China
}%

\date{\today}% It is always \today, today,

\begin{abstract}
There are more than eight hundred interest rates published in China bond market every day. Which are the benchmark interest rates that have broad influences on most interest rates is a major concern for economists. In this paper, multi-variable Granger causality test is developed and applied to construct a directed network of interest rates, whose important nodes, regarded as key interest rates, are evaluated with inverse Page Rank scores. The results indicate that some short-term interest rates have larger influences on the most key interest rates, while repo rates are the benchmark of short-term rates. It is also found that central bank bills¡¯ rates are in the core position of mid-term interest rates¡¯ network, and treasury bond rates are leading the long-term bonds rates. The evolution of benchmark interest rates is also studied from 2008 to 2014, and it's found that SHIBOR has generally become the benchmark interest rate in China. In the frequency domain we detect the properties of information flows between interest rates and the result confirms the existence of market segmentation in China bond market.
\begin{description}
\item[PACS numbers]
89.65.Gh, 89.75.Hc
%\item[Key Words]
%Causality Network, Benchmark Interest Rates, Information Flows
\end{description}
\end{abstract}

\pacs{Valid PACS appear here}% PACS, the Physics and Astronomy
%\keywords{Causality Network, Benchmark Interest Rates, Information Flows}%Use showkeys class option if keyword
\maketitle

\section{\label{sec:level1}INTRODUCTION\protect\\ }
In recent years, there has been growing interest of physicists in economic systems~\cite{mantegna1999introduction}. Among the literature using physical method to study economics, complex network provides us a visual and useful tool to learn the properties of complex systems. Complex network analysis, which can display the relations of nodes in real world in the form of network, is firstly developed to study the topology structure of some real networks like internet, movie actor collaboration network, citation networks, and so on~\cite{albert2002}. In financial market, complex network has been successfully applied in the analysis of investigating the financial contagion in the banking system and the inner structures of global stock markets based on the links and directions between the nodes~\cite{nier2007,buccheri2013,mantegna1999,tumminello2011,bech2010,kumar2012,song2011}. Complex systems are classified into different clusters according to the links between the agents, also dynamic evolutions of the system can be described with the network diagrams in different time windows. Some other scholars studied the correlations between different equities within a nation's stock market, which are helpful for investors to construct the portfolio and diversify the investment risks~\cite{buccheri2013,mantegna1999,tumminello2011,chuan2013}. What's more, besides analysis of the hidden relations between the agents, network of corporations' ownership can be constructed if company A is the shareholder of company B, then the extracted networks with out-degrees or in-degree analysis will tell us who are the important shareholders for the sample corporations~\cite{glattfelder2009}.\\

Stock market has caught many physicists' attention. According to preliminary statistics (based on data from www.webofscience.com), there are around 3400 papers on financial physics, in which nearly one fifth focus on stock market. But bond market, as an important part of financial markets likewise, whose inner relations among bond rates (also known as interest rates) receive scant attention in most econo-physics literatures. Interest rates play an important role in the investment decision-making process and asset allocation~\cite{fabozzi2003}, and many papers by economists have suggested that there exist tight links between interest rates and economic indicators (output growth, inflation, consumption, investment, labor, and stock price, etc.)~\cite{ang2006,carr1972,taylor1999,estrella2005,bansal2012,rudebusch2008,gali2014}
However, few economists or physicists paid attention to the relations between interest rates in the bond market. Various interest rates are fluctuating in their own speeds, whether there are benchmark interest rates which have broad influences on the other interest rates? And which rates are the benchmark interest rates? In this paper, we make an attempt to detect the benchmark interest rates in China bond market with the complex network method and to study the topological structure of the constructed network. The determination of benchmark rate will be helpful for conducting the transmission of monetary policy, and also the financial product valuation, as well as the process of interest rate marketization~\cite{duan2014,Wen2005}.\\

The benchmark interest rates are different around the world. In U.S., federal funds rate or three month bill rate is usually regarded as the benchmark rate~\cite{rudebusch1995federal,gutierrez2012competing,duan2014,
hamilton2002,kuttner2001monetary} . While in U.K., one or three month LIBOR (London Interbank Offered Rate ) is viewed as the core rate ~\cite{duffie2015reforming,duffie2014benchmarks}. Ref.~\cite{dunne2002}studied the interest rates in euro zone, and found a more complex structure between interest rates. The benchmark interest rate is not simply the lowest rate in a specific period but the rate which is the pricing benchmark for most other rates. Until now, there is no benchmark interest rate accepted universally in China. In earlier periods, the interbank offered rates or the repo rates are tested to be the benchmark of Chinese bond market ~\cite{Wen2005,dai2006,zhao2015testing}, while Shanghai Interbank Rates (SHIBOR) is deemed to be a better benchmark index in recent studies~\cite{Fang2012,wang2012}. Nearly all the literatures on the determination of benchmark interest rates in China tended to get results from qualitative analysis without quantitative studies, and focus on the limited sample of short-term interest rates only.\\
In this paper, bond market is regarded as a complex system including all the main types of bonds on key terms. There are a lot of correlations between various interest rates, which can be used to construct the structure network of interest rates. Granger causality test which has been widely applied in the field of economics, is to detect the causal relationship between two time series initially~\cite{wiener1956theory,granger1969,geweke1982measurement}, and then been extended to multivariable time series analysis~\cite{geweke1984measures,guo2008partial,barnett2014,chuan2013}.
The initial bivariate Granger causality test may have some inherent limitations. For example, if rate A has an influence on rate B, while B affects another rate C, but A does not have any influence on C. When we test the causal relationship between A and C using pairwise Granger causality test, the wrong patterns of connectivity will be reported. To avoid this, unlike earlier studies only focused on two time series, this paper uses the multivariable Granger causality test to form a directed network structure which shows the causal links and inter-relationships between nearly all the interest rates in China. Based on the topology structure of complex network, we can find a key rate in the core position, which is regarded as the benchmark interest rate. Information flows in the network are also studied in the frequency domain. And we find some proofs of the existence of market segmentation in Chinese bond market.\\
 The paper is organized as follows: In Sec. 2 the data sample is briefly introduced and the basic Granger causality test is outlined with an emphasis on the multivariable form. This is followed in Sec. 3. by some empirical studies to detect the benchmark interest rates in China, as well as its dynamic properties. Market segmentation and information flows are then detected in Sec. 4. Finally, we end with a conclusion in Sec.5.\\

\section{\label{sec:level1}DATA AND METHODOLOGY\protect\\ }
\subsection{\label{sec:level2}Data}
Our sample is composed of more than 800 interest rates published by China Central Depository and Clearing company (CCDC) everyday, including nearly all the bonds traded in the interbank market. Earlier studies mainly focused on short-term money market rates, like repo rates, SHIBOR, etc. In this paper we almost use the whole sample for analysis, including corporate bond rates, short-term bill rates, policy financial bond rates, and so on. The study period is from January 1, 2008 to December 31, 2014. The basic data is from \emph{Wind}. In order to guarantee the data validity, priority should be given to data cleaning process. After screening out the missing data, the time series less than 3 years and also the bonds trading inactively, there are 138 interest rates left for this study, including the main types of China bond market in short terms, middle terms and long terms. The sample data and its abbreviations are as in Table 1, in which R7d means 7-day Pledged Repo Rate, and HRCB7y is 7-year High Rank Corporate Bond rate, and so on.\\
\begin{table*}
\caption{\label{tab:table3}The sample data and its abbreviations of our reasearch}
\begin{ruledtabular}
\begin{tabular}{ccccc}
Types&Abbreviation&Terms&Sample Size\\\hline
Pledged Repo Rate&R&1d-14d&3\\
Interbank Offered Rate&IBO&1d-14d&3\\
SHIBOR&SHIBOR&1d-1y&8\\
Central Bank Bill&CBB&1d-3y&7\\
Treasury Bond&TB&1d-10y&19\\
Policy Financial Bond&PFB&1d-10y&12\\
High Rank Corporate Bond&HRCB&1d-10y&12\\
Low Rank Corporate Bond&LRCB&1d-10y&12\\
High Rank Short-term Note&HRSN&1d-5y&9\\
Low Rank Short-term Note&LRSN&1d-5y&9\\
Senior Rank Commercial Bank Financial Bond&SeCBFB&1d-10y&11\\
Subprime Rank Commercial Bank Financial Bond&SuCBFB&1d-10y&12\\
Railway Debt&RD&1d-10y&11\\
Asset Backed Security&ABS&1d-10y&10\\

\end{tabular}
\end{ruledtabular}
\end{table*}

\subsection{\label{sec:level2}Methodology}
\subsubsection{Granger Causality Test in Time Domain}
Granger Causality Test is a widely used method in economics to examine whether the two data series have Granger causal relationship~\cite{wiener1956theory,granger1969}. If the series X is helpful to reduce the prediction errors of another series Y, that is, the past of X conveys some information about the future of Y, it is said that X Granger cause Y. Yet it is a very useful method, Granger causal relationship only has statistical significance. It does not necessarily tally with the ¡°causality¡± of physics or even common sense (it has been proved in Ref.~\cite{pearl2009,valdes2011,friston2013}). This paper does`t engage in this debate here, it is enough for our study in the sense of Ref.~\cite{wiener1956theory,granger1969} as just described.\\
Granger causality between two series is usually named the pairwise G-causality, or the unconditional G-causality form. Suppose there are two time series, X and Y. And people want to detect the causal relationship from Y to X. A vector autoregressive model (VAR) may be considered. In the VAR formulation, this notion is operationalized as follows: \\
The VAR(p) decomposes as
\begin{equation}
\left( \begin{array}{l}
{X_t}\\
{Y_t}
\end{array} \right) = \sum\limits_{k = 1}^p {\left( {\begin{array}{*{20}{c}}
{{A_{xx,k}}}&{{A_{xy,k}}}\\
{{A_{yx,k}}}&{{A_{yy,k}}}
\end{array}} \right)} \left( \begin{array}{l}
{X_{t - k}}\\
{Y_{t - k}}
\end{array} \right) + \left( \begin{array}{l}
{\varepsilon _{x,t}}\\
{\varepsilon _{y,t}}
\end{array} \right)%.
\end{equation}
 with the residual covariance matrix as
 \begin{equation}
\Sigma  \equiv {\mathop{\rm cov}} \left( \begin{array}{l}
{\varepsilon _{x,t}}\\
{\varepsilon _{y,t}}
\end{array} \right) = \left( {\begin{array}{*{20}{c}}
{\Sigma _{xx}}&{\Sigma _{xy}}\\
{\Sigma _{yx}}&{\Sigma _{yy}}
\end{array}} \right)
\end{equation}
The VAR model can be estimated with likelihood function or the Levinson, Wiggins, Robinson(LWR) algorithm, while order of model(p) can be determined with the Bayesian Information Criterion(BIC)~\cite{barnett2014}.\\
The x-component of the regression (1) is
\begin{equation}
{X_t} = \sum\limits_{k = 1}^p {{A_{xx,k}} \cdot {X_{t - k}}}  + \sum\limits_{k = 1}^p {{A_{xy,k}} \cdot {Y_{t - k}}}  + {\varepsilon _{x,t}}
\end{equation}
Given its own past, the dependence of X on the past of Y is encapsulated in the coefficients $A_{xy,k}$; Particularly, there is no conditional dependence of X on the past of Y if the following null hypothesis is true:
\begin{equation}
A_{xy,1}=A_{xy,2}=...=A_{xy,p}=0
\end{equation}
Then a reduced form is written omitting the past of Y:
\begin{equation}
{X_t} = \sum\limits_{k = 1}^p {{A_{xx,k}} \cdot {X_{t - k}}}   + {\varepsilon _{x,t}'}
\end{equation}
Maximum Likelihood theory~\cite{edwards1992} furnishes a natural framework for the analysis of parametric data modelling. This motivates the definition of the Granger causality statistic.
\begin{equation}
 {\mathfrak{F}_{Y \to X}} \equiv \ln \frac{{|\Sigma {'_{xx}}|}}{{|{\Sigma _{xx}}|}}
\end{equation}
Where ${\Sigma _{xx}} = {\mathop{\rm cov}} ({\varepsilon _{x,t}})$ and $\Sigma {'_{xx}} = {\mathop{\rm cov}} (\varepsilon {'_{x,t}})$ are the residual covariance matrices of the VAR models (3) and (4) respectively. In statistic sense, The value of ${\Sigma _{xx}}$ measures the autoregressive prediction errors of X based on its own previous values, whereas the value of $\Sigma {'_{xx}}$ represents the prediction errors of X based on the previous values of both X and Y. From this perspective the statistic $\mathfrak{F}_{Y \to X}$ thus quantifies the reduction in the prediction errors when the past of the variable Y is included. As ${\Sigma }_{xx}$ is always less than ${\Sigma }_{xx}'$, it is clear that $\mathfrak{F}=0$ when there is no causal relation and $ \mathfrak{F}>0 $ when there is.\\

When there are another set of series, Z say, a pairwise analysis can also be performed one by one but a spurious causality may be reported. For instance, if there is no direct causal influence from Y to X, but there are dependencies of X and Y on Z (possibly lagged), then a spurious causal between X and Y may be reported. This paper brings forward a generalized Granger causality test of two rate series conditional on the others, then the partial effect of Z may be eliminated on causal inference~\cite{geweke1984measures}.\\
To avoid excessive mathematical complexity we develop the analysis process for three time series. The framework can be generalized to three sets of time series.
Assume Z as another variable, we wish to eliminate any joint effect of Z on the inference of the G-causality from Y to X. Again we may consider a similar VAR(p) model as (1)and (2) but with an another variable Z, and get the full and reduced form of X:
\begin{eqnarray}
{X_t} = &\sum\limits_{k = 1}^p {{A_{xx,k}} \cdot {X_{t - k}}}  + \sum\limits_{k = 1}^p {{A_{xy,k}} \cdot {Y_{t - k}}}\nonumber\\
  &+ \sum\limits_{k = 1}^p {{A_{xz,k}} \cdot {Z_{t - k}}}  + {\varepsilon _{x,t}}
\end{eqnarray}
\begin{equation}
  {X_t} = \sum\limits_{k = 1}^p {A{'_{xx,k}} \cdot {X_{t - k}}}  + \sum\limits_{k = 1}^p {A{'_{xz,k}} \cdot {Z_{t - k}}}  + \varepsilon {'_{x,t}}
\end{equation}
The null hypothesis to test the causal relationship is still (3) and the statistic is again written as:
\begin{equation}
 {\mathfrak{F}_{Y \to X| Z}} \equiv \ln \frac{{|\Sigma {'_{xx}}|}}{{|{\Sigma _{xx}}|}}
\end{equation}
which measures the degree to which adding the past of Y helps to predict X, compared with X is already predicted by its own past values and the past of Z.

\subsubsection{Granger Causality Test in Frequency Domain}
Not only in time domain, Granger causality test can also be decomposed in a natural way in frequency domain~\cite{geweke1982measurement,geweke1984measures}.Results of causality in time domain previously introduced can actually be seen as an average over all frequencies of the spectral causality. With Flourier transform procedure, equation(1) is rewritten into a spectral form.
\begin{equation}
\left( \begin{array}{l}
{X(\lambda)}\\
{Y(\lambda)}
\end{array} \right) =  {\left( {\begin{array}{*{20}{c}}
{{H_{xx}(\lambda)}}&{{H_{xy}(\lambda)}}\\
{{H_{yx}(\lambda)}}&{{H_{yy}(\lambda)}}
\end{array}} \right)} \left( \begin{array}{l}
{E_{x}(\lambda)}\\
{E_{y}(\lambda)}
\end{array} \right)
\end{equation}
where the transfer function $H(\lambda)$ is
\begin{equation}
H(\lambda )={(1-\sum_{k=1}^{p}{A}_{k}{e}^{-ik\lambda })}^{-1}
\end{equation}

After proper ensemble averaging we have the spectral matrix( which is also called cross-power spectral density, CPSD):
\begin{equation}
S(\lambda)=H(\lambda)\Sigma {H(\lambda)}^{*}
\end{equation}
Derived from (12), the CPSD of the sub-process X is written as:
\begin{eqnarray}
 S_{xx}(\lambda)=&H_{xx}(\lambda)\Sigma_{xx} H_{xx}(\lambda)^*+2\mathfrak{Re}\{H_{xx}(\lambda)\Sigma_{xy} H_{xx}(\lambda)^*\}\nonumber\\
 &+H_{xy}(\lambda)\Sigma_{yy} H_{xy}(\lambda)^*
\end{eqnarray}
It is instructive to consider the case there is no instantaneous causality, where $\sigma_{xy}=0$. We can always get this to be true using a linear transformation introduced by Geweke~\cite{geweke1982measurement}. Thus equation(13) can be written into a simpler form:
\begin{equation}
 S_{xx}(\lambda)=H_{xx}(\lambda)\Sigma_{xx} H_{xx}(\lambda)^*+H_{xy}(\lambda)\Sigma_{yy} H_{xy}(\lambda)^*
\end{equation}
whereby the CPSD of X splits into an ``intrinsic'' term and a ``causal'' term. Thus the statistic is constructed as
\begin{equation}
 \mathfrak{f}_{Y\rightarrow X}(\lambda)\equiv ln(\frac {\mid S_{xx}(\lambda)\mid}{\mid S_{xx}(\lambda)-H_{xy}(\lambda)\Sigma_{yy} H_{xy}(\lambda)^*\mid})
\end{equation}
The conditional case is less straightforward. Yet we can easily transfer the question into an unconditonal form and solve it as previously stated~\cite{geweke1984measures}. The conditional Granger causality test procedure both in time domain and in frequency domain has been integrated in a Matlab toolbox developed by Barnett and Seth~\cite{barnett2014}.\\

In this paper, we apply the conditional G-causality test process mentioned above between every interest rate in our sample conditional on all other rates. The G-causality structure between interest rates can be described with a graphical method, where the node represents one rate, and the directed edge linking two rates is specified by the G-causality relationship which has passed the significance test. In details, if Y Granger causes X conditional on all other rates, then we¡¯ll see a directed edge from Y to X.\\

\subsubsection{Inverse PageRank£¨CheiRank£©}
The Page Rank algorithm, developed by Larry Page and Sergey Brin in 1998, is a widely used sorting algorithm for web pages~\cite{Page1999}. It is the backbone of Google's search engine operations which provide the most useful service available on the internet. Page Rank is a way to measure the importance of website pages by counting the number and quality of links to a page. The underlying assumption is that more important websites are likely to receive more links from other websites. But if the website has some links targeting to other websites, the score may be reduced as well.\\
Imagine a web surfer who jumps from one web page to another, choosing with uniform probability (named damping factor) to follow at each step. When we consider the web in a directed network, the d is the probability (damping factor) in which we jump from one node to an out-linking one. Let N be the number of nodes in this network, $p_i$ be the node what we want to give a score, $p_j$ be the nodes what $p_i$ links to, and $L(p_i)$ be the set of pages which links outbound page $p_i$. For pages which have no out-links we add links to all pages in the network. In other word, this no-linking page has out-links redistributed uniformly to all pages. So the probability the surfer is on page $p_i$ at some point in time is given by the formula:

\begin{equation}
PR({p_i}) = \frac{{(1 - d)}}{N} + d\sum\limits_{j \in L({p_i})} {\frac{{PR({p_j})}}{{|L({p_j})|}}}
\end{equation}

$PR({p_i})$ is the probability, also what we called the PageRank score of node $p_i$. The calculation will continue until the $PR({p_i})$ converges. The damping factor d is usually set to be 0.85 (see Ref.~\cite{Page1999}).\\
In our research, the G-causality relation network is applied to find the interest rates which have influences on others. In word of network, that is to find a node with more out-degrees and fewer in-degrees. It's exactly opposite from the view of PageRank algorithm, which is used to find a node with more in-degrees and fewer out-degrees. So it needs to swap the links' directions and invert the adjacency matrix to the opposite. Then $L(p_i)$ is the set of nodes which links inbound $p_i$ . In this inverse PageRank way(or CheiRank introduced by Chepelianski~\cite{chepelianskii2010towards}), this paper evaluates the importance of interest rate nodes. \\
The important nodes evaluated by out-degrees and CheiRank exhibit some differences more or less in most situations.
Usually, nodes with large out-degrees may also have large in-degrees(as shown in Fig.1), which in fact, reduce the importance of the nodes. So we will take the in-degree into account, and the inverse Page Rank valuation is more reliable in this study which eliminates some nodes wrongly chosen (like SeCBFB8y in our research)than out-degree valuation .\\
\begin{figure}[b]
\includegraphics[scale=0.5]{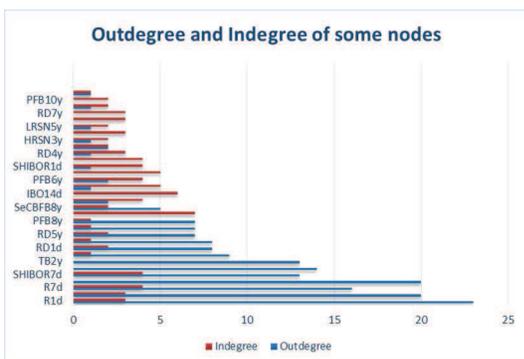}% Here is how to import EPS art
\caption{\label{fig:epsart} Out-degree and in-degree of some nodes.}
\end{figure}

\section{\label{sec:level1}BENCHMARK INTEREST RATES IN CHINA\protect\\ }
Application with Granger causality test variables should be at least weakly stationary. Also, since most rates have the same fluctuation trends, the relationships between interest rates are not correlated with the fluctuation trends. Thus removing the trend is necessary by subtracting its own moving average of 100 days (100ma) from each interest rate. Other choices also have been tried to subtract 60ma, 80ma, and 120ma from the original data, while the 100ma is the best de-trending method to make the data to be stationary and grasp the fluctuation feature as well. In this paper, a time series is stationary, only means that it has passed the stationary test ---ADF test and KPSS test.\\
After the de-trending process, all the interest rate series are stationary at 5\% significance. Conditional Granger causality test is applied to analyze the causal relationships for every two interest rates of the sample conditioning on others. Based on the analysis with Conditional Granger causality test, G-causality network of a sub-sample containing main types and terms of interest rates is constructed as in Fig 2.\\
\begin{figure}[b]
\includegraphics[scale=0.25]{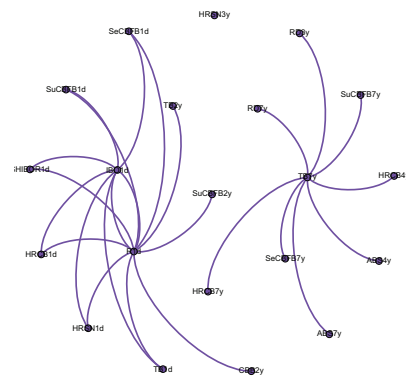}% Here is how to import EPS art
\caption{\label{fig:epsart} G-causality network of a sub-sample.}
\end{figure}
The Granger causal pattern is apparently divided into two parts. On the left side, overnight inter-bank offered rate(IBO1d) influences nearly all the short-term interest rates, while overnight repo rate(R1d) has influence on some mid-term rates. On the right side, 7-year treasury bond rate(TB7y) is in the core position of long-term interest rates and has causal influence on all long-term rates. \\
Apparently, causal network of interest rates in short-term and long-term are segmented mainly determined by the maturities, independent of each other. We also construct some networks with other sub-samples and got similar results of segmented structure as Fig.2 shown (some are divided into three or more parts). This result is in accordance with the market segmentation theory of term structure that interest rates of different terms may deliver different information of financial market~\cite{kidwell1982behavior}. The character of information flows will be discussed in the frequency domain later.\\

\subsection{\label{sec:level2}Maturities Analysis}
To be specifically, next we will divide sample of interest rates into three groups according to their maturities, and study the G-causality network structure for each group. The rates of maturities no longer than one year are in short-term group, and the maturities between one and five years are in mid-term group, and maturities longer than five years are in long-term group.
The size of the nodes is dependent on their CheiRank score, that is, the more influences an interest rate has on others, the larger this node will be.\\
Fig.3 shows the network of each group. We can see clearly there are few nodes with large CheiRank scores while others are much smaller. There exists an interest rate which influences most of rates within the group. Fig.3(a) shows the result of short-term group. We can see clearly that there are three nodes significantly larger than others which represent R1d, IBO1d, and R7d. All of the three are money market rates, at which financial institutions call loans to address short term financing needs. Commercial banks and corporations pledge securities(mainly treasury bonds) for short-term financing needs and repurchase the securities later. Fig. 3(b) shows the G-causality network of mid-term group and also its CheiRank score distribution. The biggest node according to the CheiRank score is 2-year central bank bill rate(CBB2y). For lack of short or mid term treasury notes in China, the central bank bills(CBB) are often seen as the symbol of short term national credit. Meanwhile, CBB are basic tools for monetary policy, so the change of CBB rates are often regarded as a kind of policy signal. In the long-term group as shown in Fig 3(c), 7-year treasury bond rate (TB7y) which has the largest CheiRank score, is at the key position affecting most of long term interest rates. This result is also convincing since 7-year treasury bonds are most actively traded treasuries in China.\\

\begin{figure*}
\centering
\subfigure[short-term]{\includegraphics[scale=0.2]{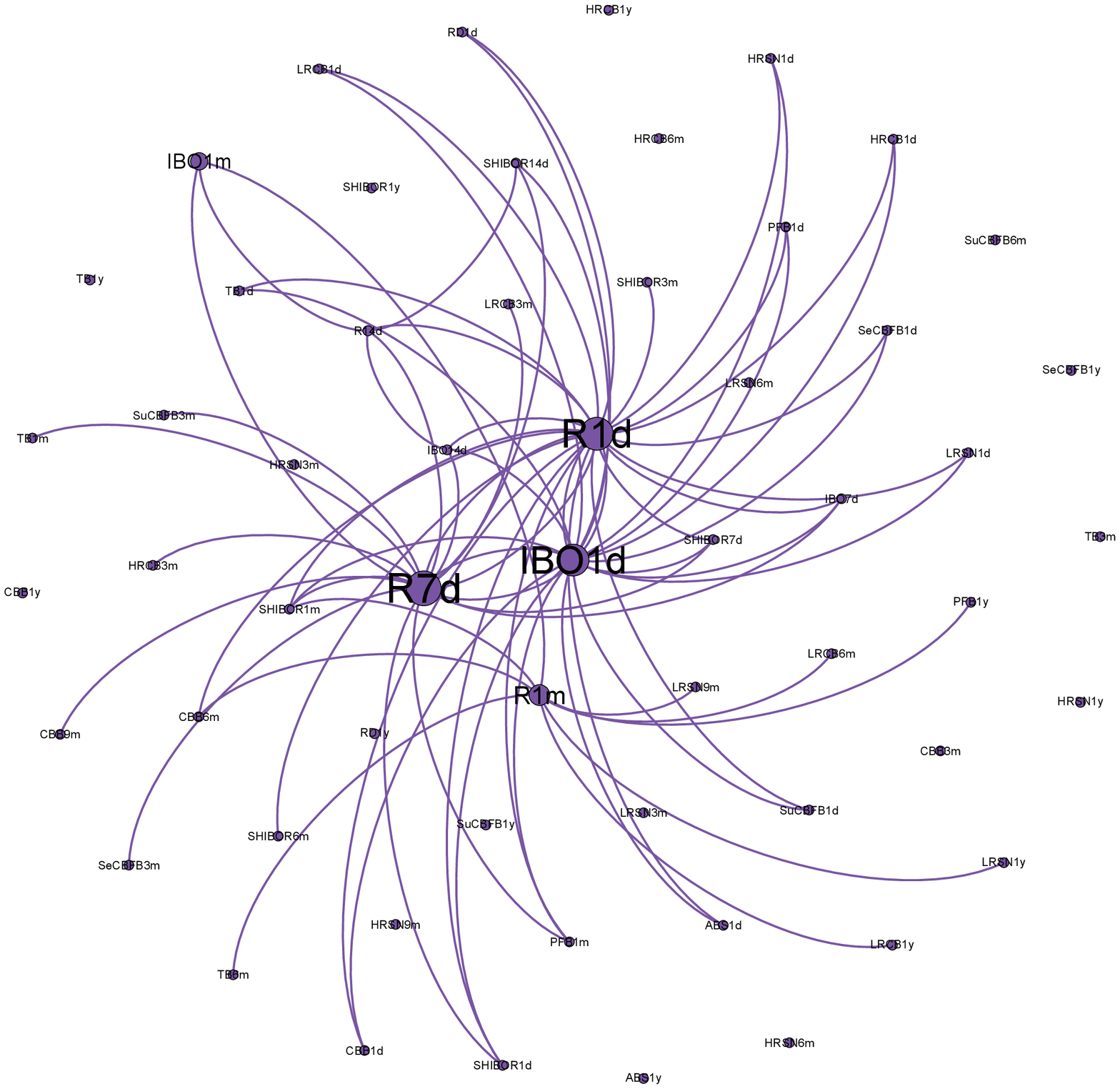}}% Here is how to import EPS art
\hspace{0.3in}
\subfigure[mid-term]{\includegraphics[scale=0.2]{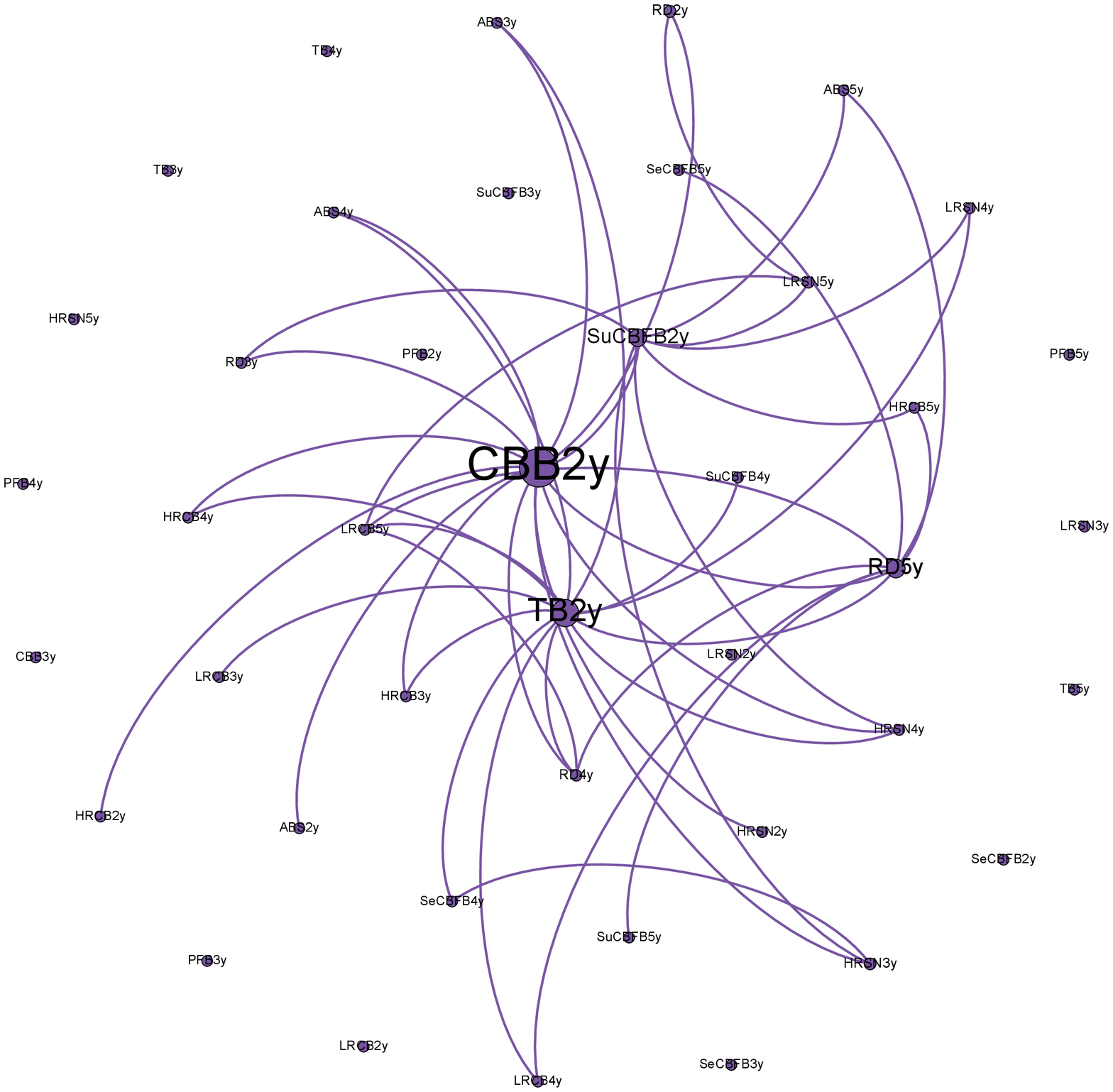}}
\hspace{0.3in}
\subfigure[long-term]{\includegraphics[scale=0.2]{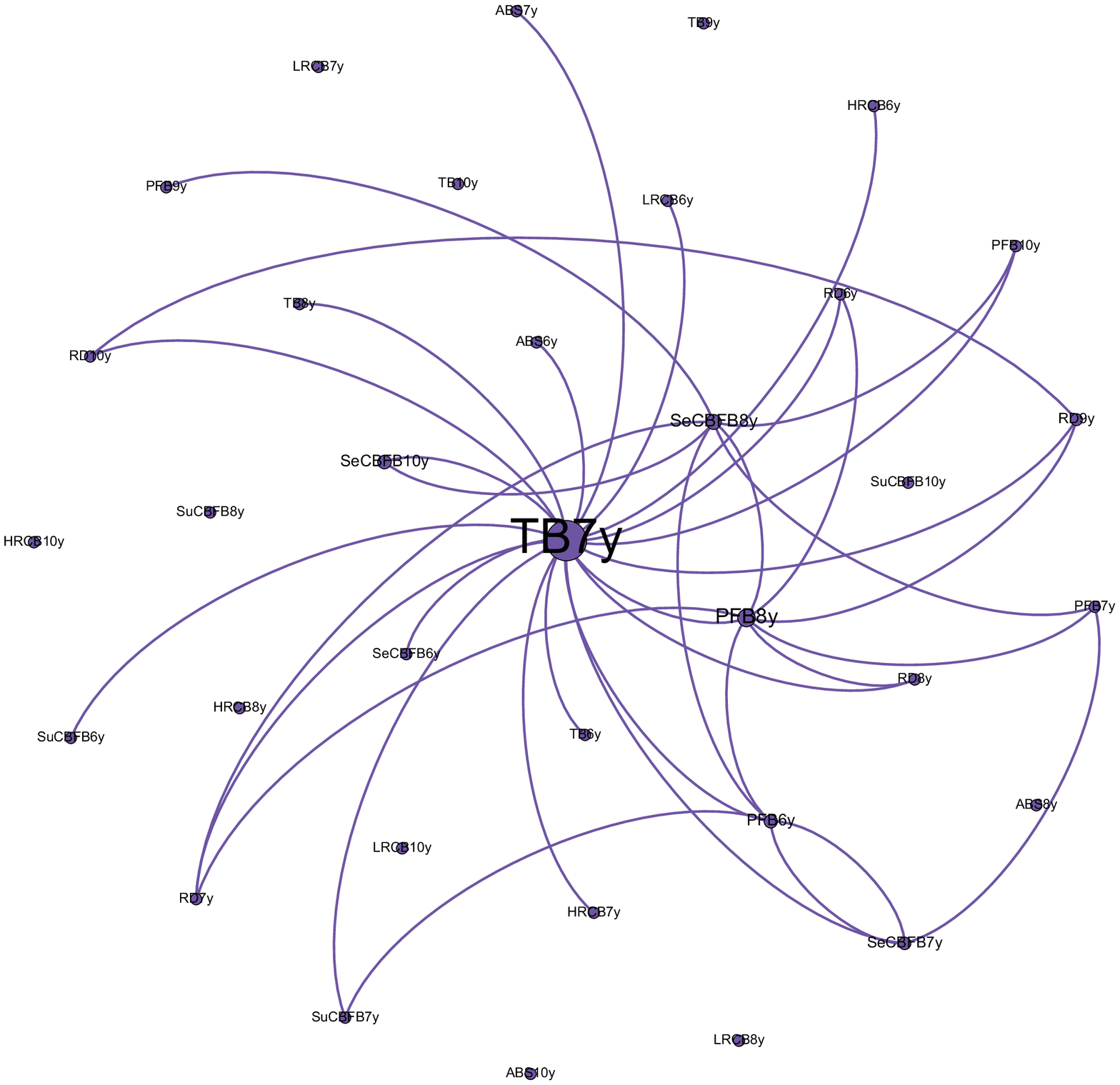}}
\caption{\label{fig:wide}Granger causality network}
\end{figure*}

Summarize the above empirical study briefly, repo rates are absolutely in the center of short-term rates, which have influences on nearly all the money market interest rates. Meanwhile IBO rates also play strong guiding role. 2-year central bank bill rate (CBB2y) plays the leading role in the mid-term group. And TB7y alone is in the center of long-term interest rates.\\
According to the analysis of this paper, interest rates in core positions like money market rates, central bank bill rates and treasury bond rates are all relatively risk free compared to the corporate bonds, financial bonds, and other credit bonds. This result implies that the valuation of credit bonds should refer to the risk-free rates. Among the risk-free rates, R1d, CBB2y, and TB7y are at the benchmark position, which are the benchmark interest rates in China bond market. One possible explanation is that repo rates reflect the change of short term money supply in the market, the CBB rates mainly transfer the signal of policy factor which is a mid-term impact, while the long-term treasury bond rates reflect the change of economic fundamentals, which is in accordance with the conclusions of Ref.~\cite{dong2011} summarized by practice and observation. \\

\subsection{\label{sec:level2}Evolution of Benchmark Interest Rates}
We've found that the repo rates, central bank bill rates, and treasury bond rates represent the properties of benchmark interest rates in different terms. While SHIBOR, interbank offered rates, railway debt rates and policy financial bond rates have certain characteristics of benchmark more or less. In this Section, we will study the evolution of benchmark interest rates on time scale. The corporate bonds and short-term notes have not represented any benchmark properties in the above study, so we remove the two categories to meet the freedom degree requirement of the model.\\
We divide the full time series into three rolling windows: stage 1 (Jan. 1, 2008 to Dec.31, 2010), stage 2 (Jan. 1, 2010 to Dec. 31, 2012), and stage 3 (Jan. 1, 2012 to Dec.31, 2014), all of which have 36 months. The result in each stage is shown in Fig. 4, and the nodes with higher CheiRank scores in each window are shown in the following list.\\
Fig 4(a) shows the G-causality network of stage 1 (2008-2010). Repo rates and the IBO rates get the largest CheiRank score, which are in the core position. Since we have deleted the corporate bonds, and thus most of the left are short term bonds, that's why the nodes of treasury bonds do not look as large as before. Granger causality structure among interest rates in stage 2(2010-2012) is shown in Fig. 4(b). Repo rates are still in the center of network, and SHIBOR has begun to reflect the guidance role. In stage 3(2012-2014), SHIBOR has already taken the place of repo rates to be the new benchmark of bond market. Most of the nodes with largest Page Rank score are SHIBOR in different terms.\\
\begin{figure*}
\centering
\subfigure[080910]{\includegraphics[scale=0.2]{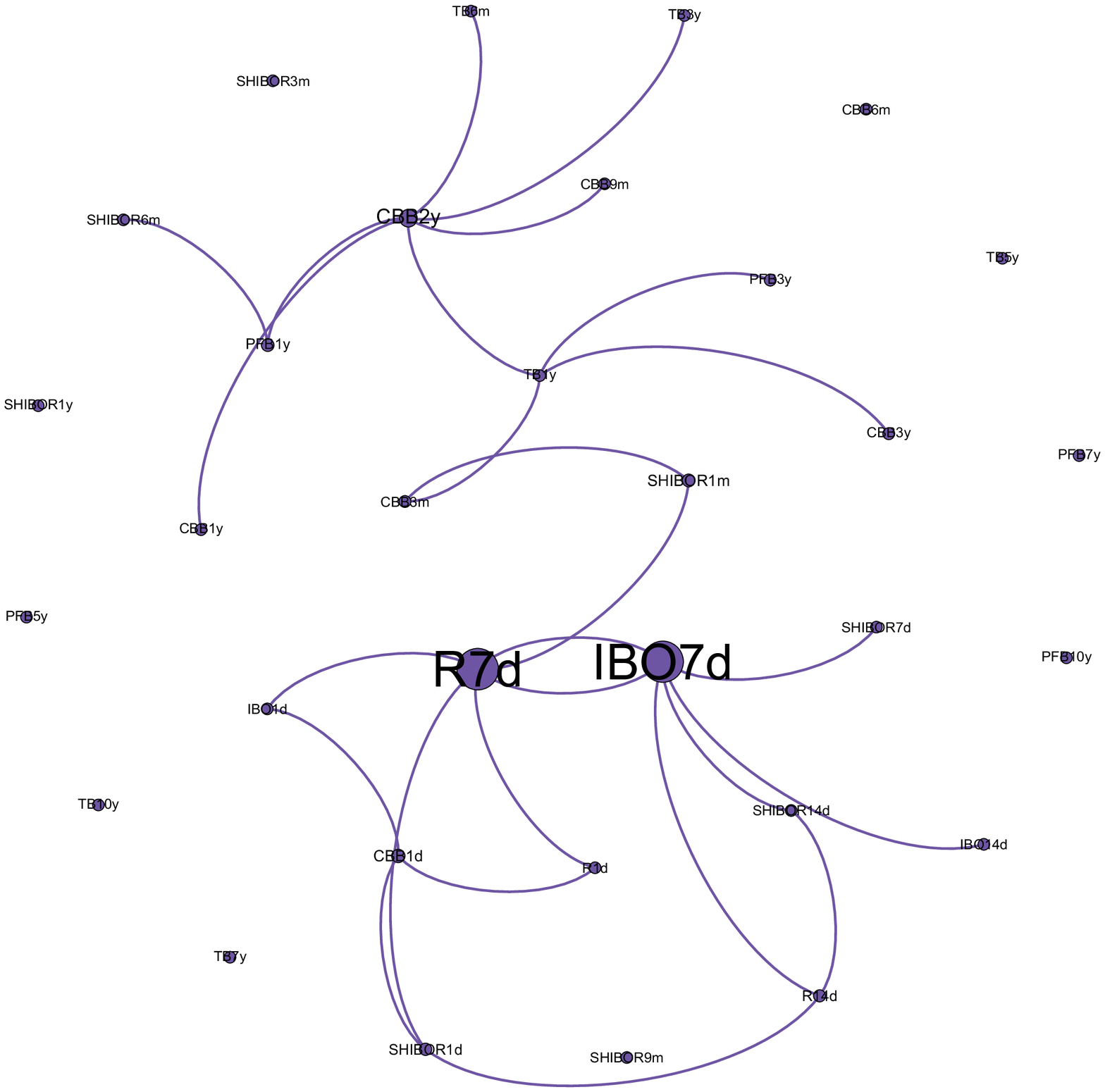}}% Here is how to import EPS art
\hspace{0.3in}
\subfigure[101112]{\includegraphics[scale=0.2]{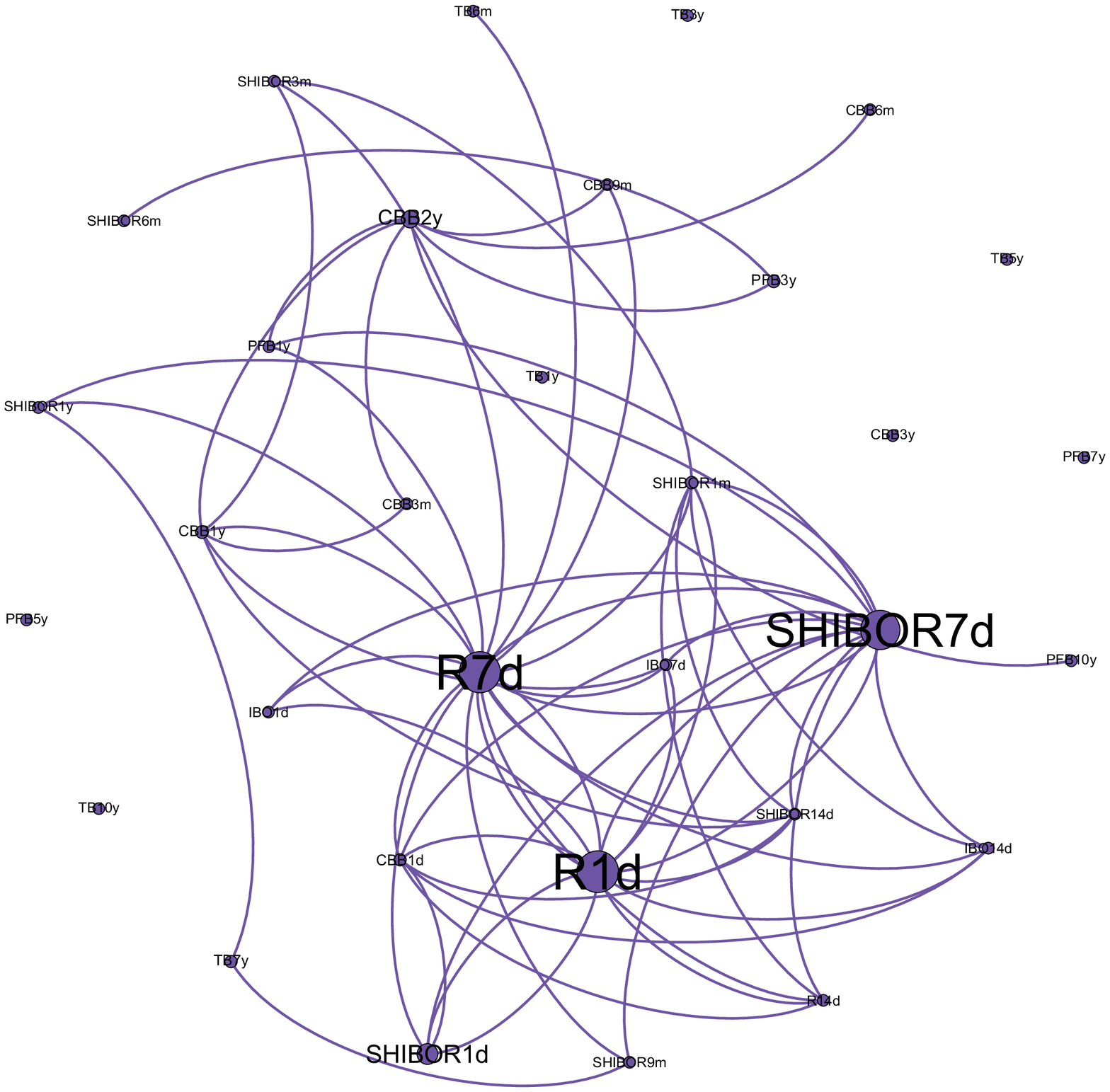}}
\hspace{0.3in}
\subfigure[121314]{\includegraphics[scale=0.2]{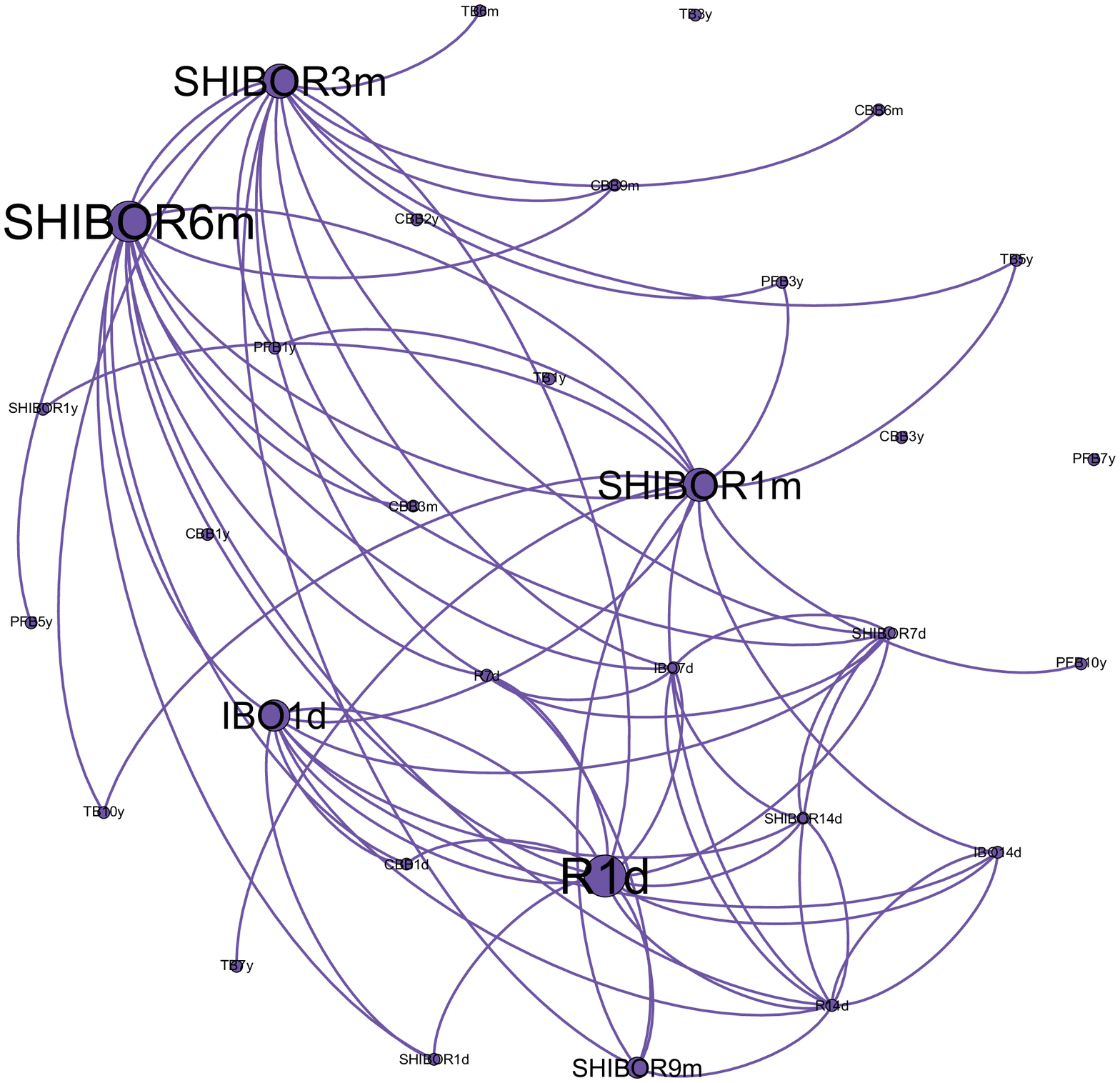}}
\caption{\label{fig:wide}Evolution of benchmark interest rates.}
\end{figure*}
Shanghai interbank offered rate, known as SHIBOR, is put forward in 2007 by central bank of China. SHIBOR is modelled on LIBOR, quoted by several high credit ranking banks everyday, and has been cultured as benchmark of Chinese bond market since the beginning. From the perspective of evolution of benchmark interest rates, we can easily see the SHIBOR has been in the core position gradually.

\section{\label{sec:level1}INFORMATION FLOWS BETWEEN RATES\protect\\ }
Edges in network are constructed by G-causality test in time domain, which is a kind of average over all frequencies of spectral causality. Thus the networks in time domain have covered up some characteristics of information flows in frequency domain. As we mentioned previously, benchmark interest rates in different maturities transfer different signals, either short-term money supply, or policy signals, or economic fundamentals. In this section, we'll construct a spectral causality network and detect the frequency features of information flows.\\
Reconsider the sub-sample shown in Fig.1, three interest rates are in the key positions to influence different rates separately according to their maturities. We construct the VAR model again and calculate the spectral causality statistic described in Eq.(16). There are three kinds of frequency spectrum as shown in Fig.5. The blue solid line represents the influence from R1d to SN1d. The influence mainly focuses on high frequency, and the information flows between R1d and SN1d is high frequent. While the impact from TB7y to CB7y(represented by the gray dotted line) is mainly in low frequency and the information flows between R1d to TB1d fall mainly in medium frequency just as the dot dash orange line shown. We then calculate all the exact frequencies where the maximum power falls in. The result is shown in Fig.6. It's clearly there are three kinds of information flows in the network. The correlation structure among interest rates are closely related with their maturities. A more intuitive graph is shown in Fig.7, where the color of edge represents the frequency of information flow between the linking nodes. The color from blue to red represents the frequency from high to low. Information flows between short and mid term interest rates are high frequent while in the right part of Fig.7, information flows in long term rates are rather low frequent. Information flows between bonds are different according to their maturities. This result confirmed the segmentation structure in China bond market again.\\

\begin{figure}[b]
\includegraphics[scale=0.6]{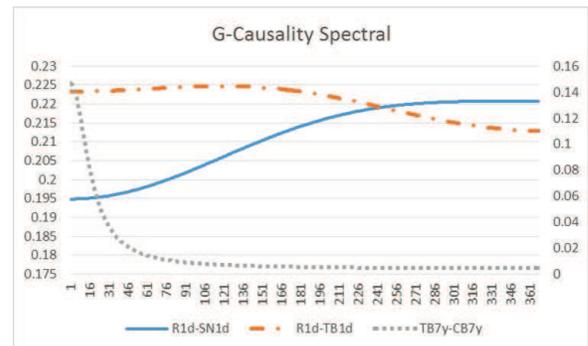}% Here is how to import EPS art
\caption{\label{fig:epsart} G-causality spectral}
\end{figure}
\begin{figure}[b]
\includegraphics[scale=0.6]{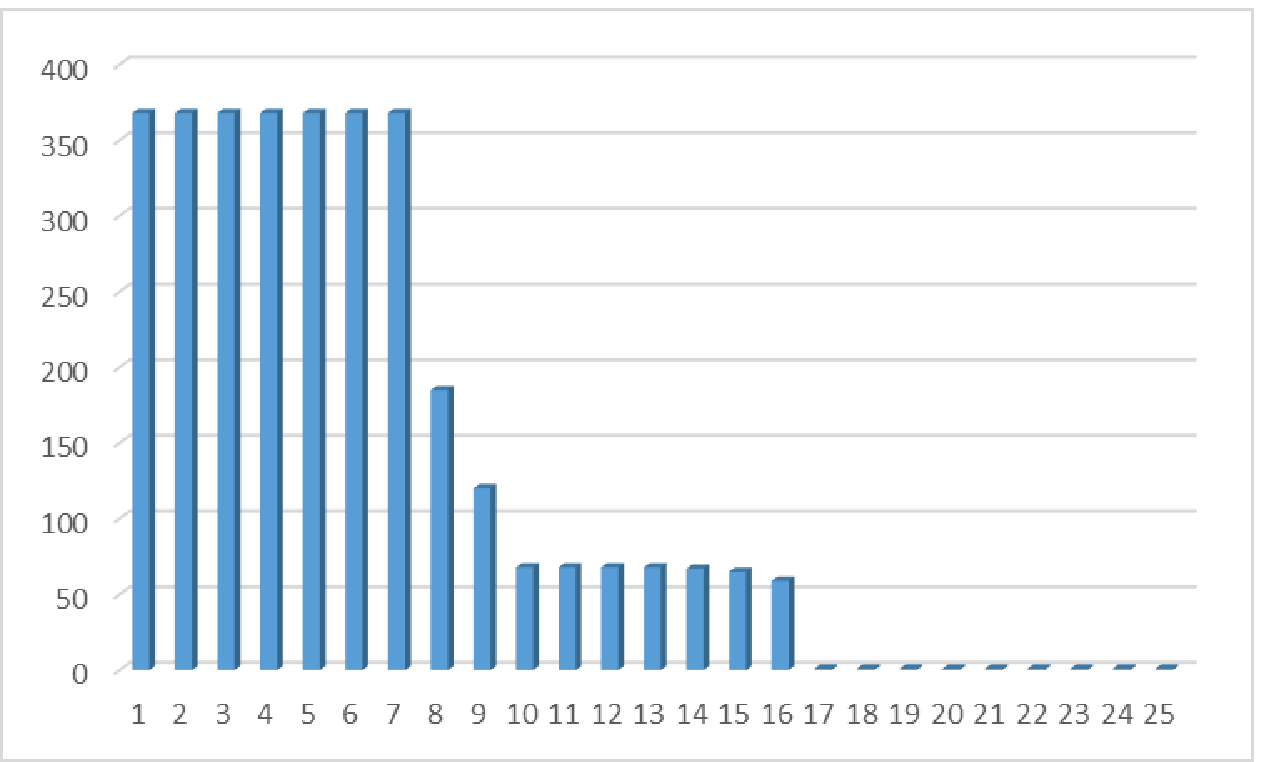}% Here is how to import EPS art
\caption{\label{fig:epsart} Frequency of information flows}
\end{figure}
\begin{figure}[b]
\includegraphics[scale=0.3]{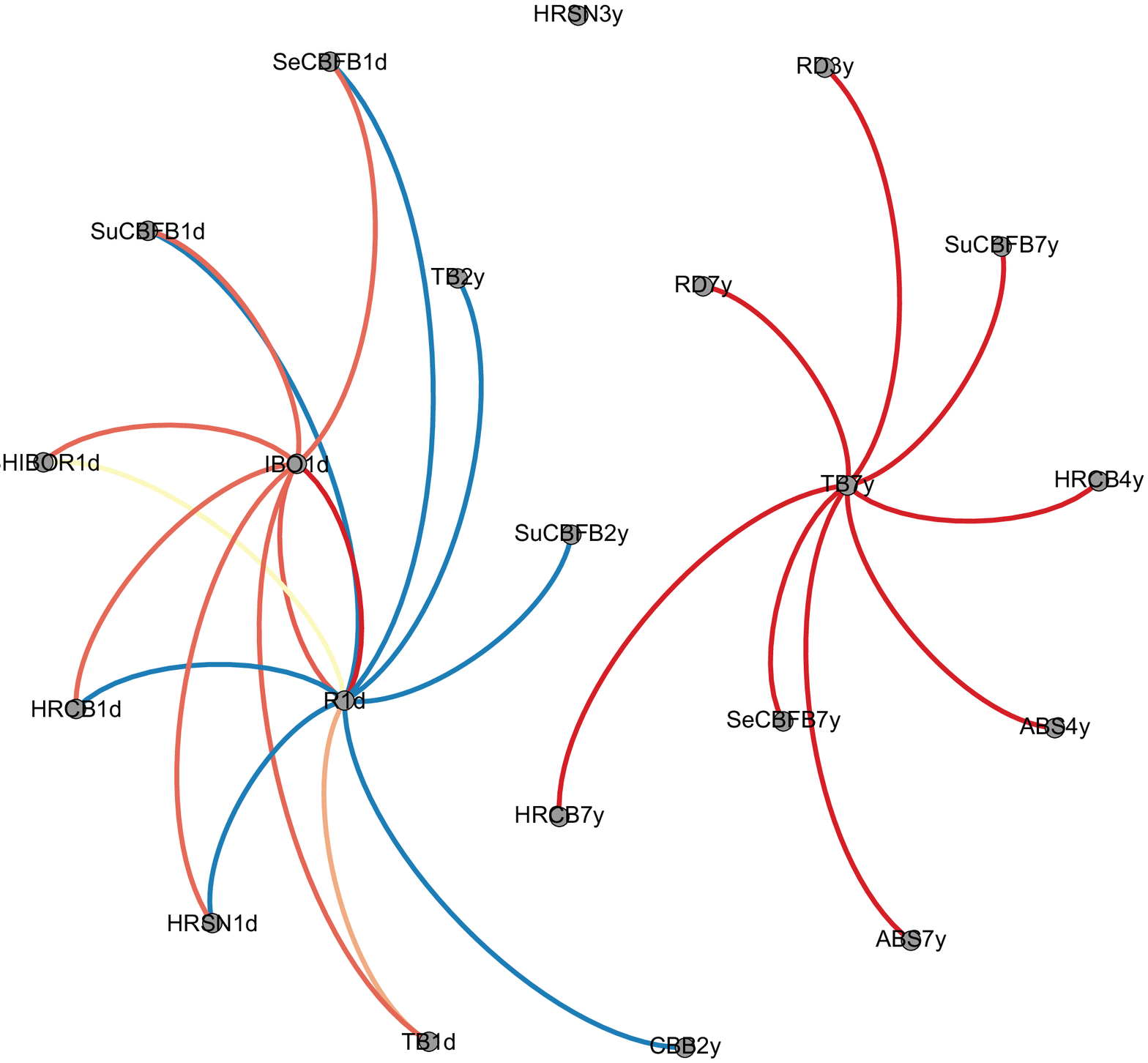}% Here is how to import EPS art
\caption{\label{fig:epsart} Information flows in network}
\end{figure}
Market segmentation theory suggests that securities can be subdivided into distinct groups according to maturities with a lack of substitution between groups exists for investors~\cite{benson1981systematic,kidwell1983market}. Once the theory put forward, evidences of market segmentation are found in municipal bond market, treasury bond market, and so on~\cite{kidwell1983market,simon1991segmentation,hendershott1978impact}. Most of the empirical studies focused on American bond market, while few evidence of market segmentation existence in China was found. Our study covers the void of research by focusing on China bond market, and we find evidence of market segmentation in China. Furthermore, earlier studies traditionally built a regression model using the bond supply and demand factors to detect the existence of market segmentation indirectly~\cite{kidwell1983market,simon1991segmentation}. By the G-causality method, this study detects the information flows between bonds directly both in time and frequency domain.

\section{\label{sec:level1}CONCLUSION\protect\\ }
This paper explores the network properties of the conditional Granger causality structure in China bond market based on a relatively large sample of interest rates and all conclusions drawn from the network analysis are in accordance with economic significance. We calculated the network structure on different terms of rates, and evaluated the importance of nodes with CheiRank score. It's found that the repo rates and IBO rates are in the core position of short term rates, and influence nearly all short term rates. The central bank bill rates have influences on mid-term rates, and the treasury bond rates influence long term rates. We then studied the evolution of benchmark interest rates from 2008 to 2014, and found that SHIBOR had increasingly been accepted by the market and becoming the benchmark of bond market. Furthermore, Using G-causality method to detect the information flows both in time and frequency domain, we also find some evidence of the existence of market segmentation in China bond market.

\begin{acknowledgments}
We thank for anonymous referees¡¯ suggestions and the financial support by National Science Foundation of China (No. 71573243) and Youth Innovation Promotion Association of Chinese Academy of Sciences. We are also grateful for the support by the Open Project of Key Laboratory of Big Data Mining and Knowledge Management of Chinese Academy of Sciences.
\end{acknowledgments}

\bibliography{benchmark_interest_rates}% Produces the bibliography via BibTeX.
\end{document}